\documentclass{jfm}

\usepackage{amsmath,amsfonts,amssymb}
\usepackage{graphicx}
\usepackage{setspace}
\usepackage{natbib}
\usepackage{hyperref}

\hypersetup{
    colorlinks = true,
    urlcolor   = blue,
    citecolor  = black,
}

\newcommand{\RomanNumeralCaps}[1]

\title{Particle Radial Distribution Function and Relative Velocity Measurement in Turbulence at Small Particle-Pair Separations}

\author{Adam Hammond\aff{1},
  Hui Meng\aff{1}\corresp{\email{huimeng@buffalo.edu}}}

\affiliation{\aff{1}Department of Mechanical and Aerospace Engineering, University at Buffalo, Buffalo, NY 14260, United States}

\newcommand{\sstar}{\includegraphics[width=0.2in]{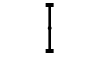}}
\newcommand{\striangle}{\includegraphics[width=0.12in]{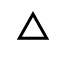}}
\newcommand{\sdash}{\includegraphics[width=0.15in]{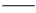}}
\newcommand{\sstardash}{\includegraphics[width=0.17in]{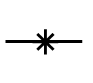}}
\newcommand{\striangledash}{\includegraphics[width=0.17in]{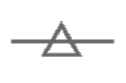}}
\newcommand{\sdasha}{\includegraphics[width=0.2in]{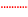}}
\newcommand{\sdashb}{\includegraphics[width=0.2in]{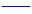}}
\newcommand{\sdashc}{\includegraphics[width=0.2in]{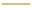}}

\begin{document}
\maketitle

\begin{abstract}
The collision rate of particles suspended in turbulent flow is critical to particle agglomeration and droplet coalescence. The collision kernel can be evaluated by the radial distribution function (RDF) and radial relative velocity (RV) between particles at small separations $r$. Previously, the smallest $r$ was limited to roughly the Kolmogorov length $\eta$ due to particle position uncertainty and image overlap. We report a new approach to measure RDF and RV near contact ($r/a\: \approx$ 2.07, $a$ particle radius) overcoming these limitations. Three-dimensional particle tracking velocimetry using four-pulse Shake-the-Box algorithm recorded short particle tracks with the interpolated midpoints registered as particle positions to avoid image overlap. This strategy further allows removal of mismatched tracks using their characteristic false RV. We measured RDF and RV in a one-meter-diameter isotropic turbulence chamber with Taylor Reynolds number $\Rey_\lambda=324$ with particles of 12-16 $\mu$m radius and Stokes number $\approx$ 0.7. While at large $r$ the measured RV agrees with the literature, when $r<20\eta$ the first moment of negative RV is 10 times higher than direct numerical simulations of non-interacting particles. Likewise, when $r>\eta$, RDF scales as $r^{-0.39}$ reflecting RDF scaling for polydisperse particles in the literature , but when $r\lessapprox\eta$ RDF scales as $r^{-6}$, yielding 1000 times higher near-contact RDF than simulations. Such extreme clustering and relative velocity enhancement can be attributed to particle-particle interactions. Uncertainty analysis substantiates the observed trends. This first-ever simultaneous RDF and RV measurement at small separations provides a clear glimpse into the clustering and relative velocities of particles in turbulence near-contact.
\end{abstract}


\section{Introduction}
\label{sect:01}  
Understanding the interaction of inertial particles dispersed in turbulence at close separations is critical to modelling particle collision rates. Turbulence drastically enhances the collision rates of water droplets in clouds \citep{RN1}, leading to a ``size gap" of particles with radii 10-50 $\mu$m \citep{RN2}. \cite{RN3} found that turbulence contributes to collision rates through particle preferential concentration and particle-pair relative velocity:
\begin{equation}
\label{eqn:01}
    k(2a)=4\pi(2a)^2g(r)|_{r=2a}\langle w_r|_{r=2a}\rangle ^-
\end{equation}
where $k(2a)$ is the collision kernel, $r$ is the interparticle separation distance measured from center to center, $a$ is the particle radius, $g(r)$ is the radial distribution function (RDF) and $w_r(r)$ is the particle-pair radial relative velocity (RV), and $\langle w_r|_{r=2a}\rangle ^-$ is the first moment of the \textit{negative} relative velocities, which will be referred to as ``inward RV" $\langle w_r|r\rangle ^-$, evaluated at contact, expressed as $\int_{-\infty}^{0}w_rp(w_r|_{r=2a})dw_r$, where $p(w_r|r)$ is the PDF of RV at a given $r$. 

These two collision-enhancing mechanisms by turbulence are strongly influenced by particle inertia measured by Stokes number $St$. For finite inertia, particles preferentially accumulate in the straining regions of the turbulent flow, enhancing the $g(r)$ at small $r$ \citep{squires1991preferential}. Inertia also disrupts the correlation of motion between particles by ejecting particles from different energetic eddies and converging them in the straining region, thereby enhancing the inward RV, $\langle w_r|r\rangle ^-$. This is known as the ``sling effect" \citep{falkovich2002acceleration,falkovich2007sling} verified experimentally by \cite{RN5}, and also termed ``path-history effect" \citep{RN6}. Both of these mechanisms contribute to higher collision rates compared to inertia-free particles. Since they depend on particle-turbulence interactions (PTI), they are relevant to $r$ scales down to approximately the Kolmogorov length $\eta$ and below. When $r$ decreases to $O(a)$, which is often $\ll\eta$, however, particle-particle interactions (PPI) become important. For example, hydrodynamic interactions (HI) arise through the disturbance of the flow field felt by one particle due to the presence of a nearby particle \citep{batchelor1972hydrodynamic}. Moreover, electrically charged particles will also experience attractive or repulsive Coulomb forces which can affect these collision statistics \citep{lu2015charged}.

\par It is extremely difficult for direct numerical simulation (DNS) to simulate PPI in turbulence due to high computational expense \citep{RN7}. Thus, simulations have been restricted to analyzing the particle collision kernel contributed solely by PTI, called the geometric collision kernel \citep{RN2}, wherein PPI are simply represented as a coefficient called the collision efficiency \citep{RN3,brunk1998turbulent} to be modelled theoretically \citep{RN8} and estimated using DNS through implementation of these models \citep{wang2008turbulent}. However, PPI may have complex influences on $g(r)$ and $\langle w_r|r\rangle ^-$ at $r\lessapprox\eta$ that are not captured in a study of the geometric collision kernel.

\par In order to accurately calculate the collision kernel, it is imperative to capture both the effects of turbulence and PPI. Physical experiments offer the advantage of retaining these physics. However, experimental measurement of the collision rate \citep{RN9} has so far been limited to direct observation of liquid droplet coalescence, wherein it is difficult to discern the mechanisms leading to the observed collision rates. Improved methods with higher resolution \citep{RN10} could improve direct collision rate observation.

\par On the other hand, the collision kernel can be estimated by approaching Eq. (\ref{eqn:01}) from the right-hand side, which nonetheless requires simultaneous RDF and RV measurements. Unfortunately, most experiments to date have lacked the spatiotemporal resolution to record particle motions at scales small enough to inform particle interactions. The first and perhaps only simultaneous experimental measurement of RDF and RV in isotropic turbulence known to us was by \cite{RN11} using 3D digital holography. While the holographic lateral spatial resolution was adequate, the limited angular aperture of early digital holograms \citep{RN12} caused excessive axial uncertainties \citep{RN13}. Furthermore, their 2-pulse nearest-neighbor particle tracking algorithm suffered severe tracking ambiguity and significant errors in RV calculations as $r$ decreased \citep{RN11}. Consequently they could not measure RDF and RV at $r\leq\eta$. However, their holographic RDF measurement (unaffected by tracking ambiguity) resulted in good comparison with DNS down to $r\approx\eta$ \citep{RN14}.

\par Thereafter, more sophisticated tracking schemes have enabled improved RV measurements with resolution down to $r\approx\eta$. \cite{RN15} studied the scaling of RV at the dissipation scales of turbulence ($r\approx\eta$) using a time-resolved 3D particle tracking velocimetry (3D-PTV) technique with 2-camera shadow imaging. \cite{RN16} studied the dependence of RV on $St$ and Taylor scale Reynolds number $\Rey_\lambda$ in isotropic turbulence (246$<\Rey_\lambda<$357) using a four-frame planar PTV system, where the smallest separation was limited to $O(\eta)$ \citep{RN17}. Meanwhile, recent RDF measurement has also overcome the $r\approx \eta$ resolution barrier. \cite{RN18} reported the first sub-Kolmogorov ($r/\eta \approx 0.2$) RDF measurement of particles in isotropic turbulence (155$<\Rey_\lambda<$314) by using a 3D-PTV technique to acquire particle positions. Their $g(r)$ clearly was drastically enhanced when $r\lessapprox \eta$, which was attributed to hydrodynamic interactions between particles. However, as $r$ went below $O(10a)$, their $g(r)$ exhibited significant scatter, possibly due to insufficient resolution.

\par Here we report the first detailed, \textit{simultaneous} measurement of RDF and RV down to near-contact, for estimation of the collision kernel with hydrodynamic interactions.

\section{Challenges in measuring RDF and RV at Small $r$}
\label{sect:02}
Measuring RDF and RV at small $r$ down to near-contact requires minimizing particle position uncertainty and particle image overlap, two factors limiting the smallest measurable $r$. In non-holographic 3D imaging, particle positioning uncertainty comes from 2D positioning uncertainties and errors in 3D mapping from multiple cameras. Using Iterative Particle Reconstruction with the recently emerged Multi-pulse Shake-the-Box algorithm brings particle position uncertainties down to 0.15 pixels \citep{RN19}, but when the pixel scale is small, this can be challenging to achieve if the experimental setup experiences any slight vibrations.

\par The second factor limiting the smallest measurable $r$, particle image overlap, is an inherent hindrance to resolving particle pairs with small separations and more difficult to mitigate than the position uncertainties. Near-contact particles may overlap in their 2D projections, leading to fused images that appear as single particles. This is exacerbated by optical diffraction of the high $f$-number lens for acquiring volumetric measurements, which enlarges the apparent particle image on the camera. 
\par To avoid the particle image-overlap problem, we have devised a novel 3D particle tracking velocimetry technique using the four-pulse Shake-the-Box (4P-STB) algorithm to accurately identify particle positions (and thus velocities) when particle separation $r$ is small. We record successive particle positions over a brief track of four instants and then identify particle position and velocity at the track midpoint for calculation of $g(r)$ and $w_r(r)$. When by chance the closest approach between two neighboring particles is near the track midpoint, this strategy allows for acquisition of particle position without image overlap, thereby drastically reducing the smallest measurable $r$. In addition, the use of the midpoint of the 4-pulse track further improves tracking accuracy by allowing the removal of mismatched particle identities (detailed in Section \ref{sect:04}). Our 4-pulse tracking approach to mitigate the barrier of near-contact image overlap is the key to our ability to cast a first-ever glimpse into the near-contact particle positions and velocities in turbulent flows for collision statistic measurements.

\section{Experimental Setup}
\label{sect:03}
\subsection{Isotropic Turbulence Flow Facility}
\begin{figure}
	\centering
	\includegraphics[scale=0.8]{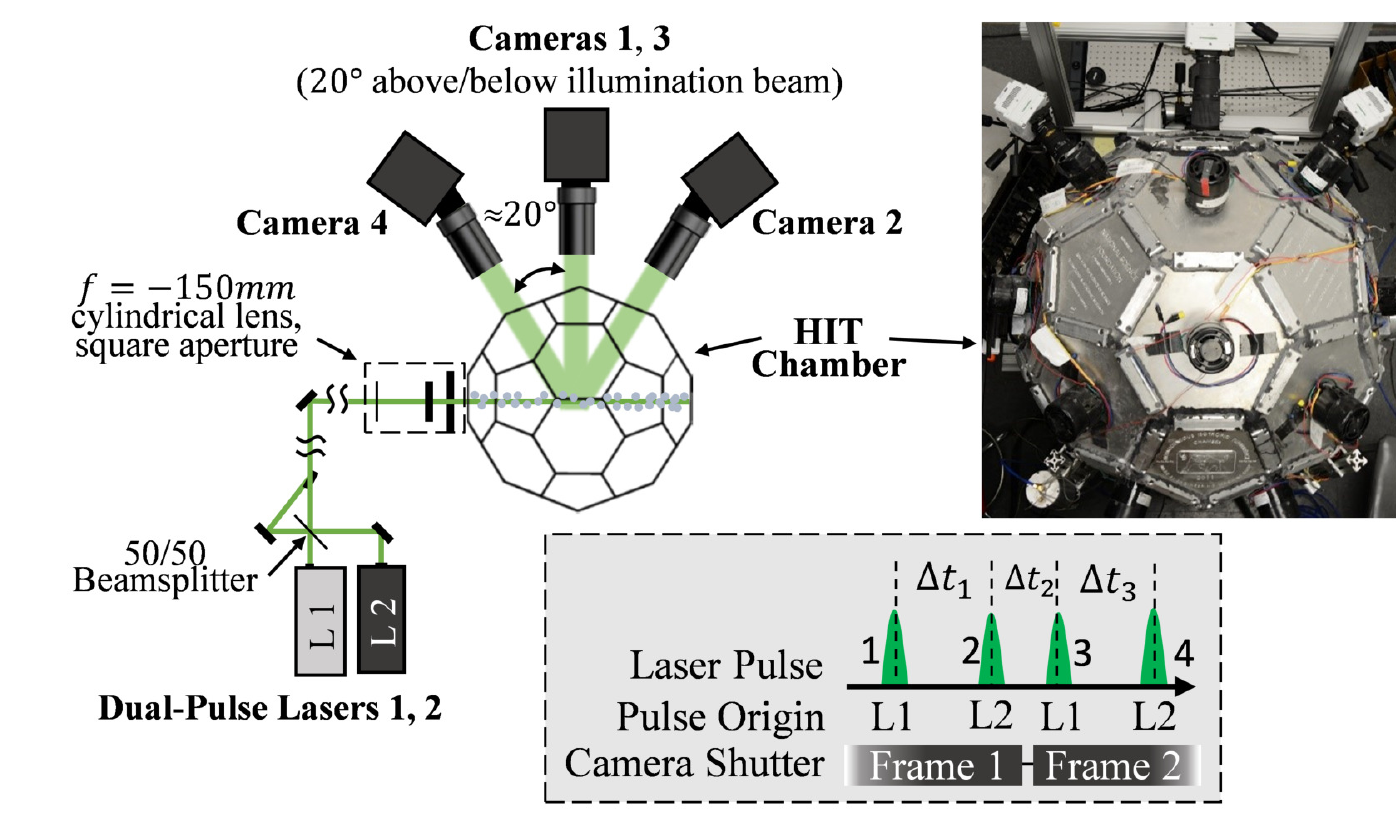}
	\caption{4P-STB system (top view) and timing.}
	\label{fig:01}
\end{figure}

We performed particle tracking in a high-Reynolds-number enclosed truncated icosahedron homogeneous isotropic turbulence (HIT) chamber (Fig. \ref{fig:01}). This 1-meter ``soccer ball" shaped facility described in \citep{RN20} is a second-generation isotropic turbulence chamber, improved from the original cubic turbulence box (8 fans with $\Rey_\lambda$ between 110 and 150) developed for our first attempt of simultaneous RDF and RV measurement \citep{RN11}. The turbulence in the HIT chamber has been completely characterized in the central isotropic region (diameter 4.8cm) by \cite{RN20}. We held the fan speed constant such that $\Rey_\lambda=324$ and used 3M K25 hollow glass spheres (3M, St. Paul, Minnesota) as particles, narrowing their diameter range to 25-32 $\mu$m using sieves following the procedure in \cite{RN17}. The average density of the sieved particles was measured using a Micromeritics Accu-Pyc II 1340 gas-displacement pycnometer. The resulting particle and flow characteristics are listed in Table \ref*{tab:01}.

\begin{table}
	\centering
	\begin{tabular}{cc}
		Particle Type                & 3M K25 hollow glass microspheres \\
		Particle Radius              & $12.5\mu \text{m} - 16\mu \text{m}$            \\
		Stokes Number                & $0.57-0.93 $                       \\
		Average Particle Density             & 0.31 g/cc                        \\
		Reynolds Number $\Rey_\lambda$ & 324                              \\
		Froude Number $Fr$ & 13.4                              \\
		Kolmogorov Length $\eta$     & $123 \mu \text{m}$                     \\
		Kolmogorov Time $\tau_k$     & $1000 \mu \text{s}$                     \\
		Kolmogorov Velocity $v_k$    & $0.13$ m/s                      
	\end{tabular}%
	\caption{Particle and flow conditions. For complete details of turbulence in the HIT chamber see \cite{RN20}.}
	\label{tab:01}
\end{table}

To reduce complexity of our experiments, we kept the electric charge and gravity effects to a minimum. To minimize triboelectric charging of the particles caused by friction with the fans and walls, the inner surfaces of the turbulence chamber were coated in conductive carbon paint and electrically grounded as described in \cite{RN17}. This helped to remove the charge on the particles. To mitigate the effect of gravity on the particles, we used fans as flow actuators in our chamber \citep{RN20}, which yielded a high Froude number $Fr=13.4$. Furthermore, due to the low density and large size of our hollow glass spheres, the gravitational settling speed (assuming Stokes drag and a quiescent flow) of our particles was $0.007 m/s$, compared to the Kolmogorov velocity of $0.13 m/s$.

To prevent any transient effects in the statistics of particle motion due to particle injection, the particles were aerosolized, then pneumatically injected into the flow facility and allowed to equilibrate over 100 large eddy turnover times ($\approx$ 30s). The particle volume fraction was kept at $\sim 2.2 \times 10^{-5}$ (equivalent to 0.002 particles per pixel, ppp) to remain well within the dilute limit.

\subsection{Optical Interrogation Setup}
\textbf{Optical configuration} - Figure 1 illustrates the 4P-STB setup for the HIT chamber. For each double-exposure double-frame recording, two dual-head Photonics Nd-YLF lasers (L1, L2) created a sequence of four independent 30mJ laser pulses fired sequentially, to record the particles over four successive instants. To direct the pulses into an interrogation volume in the chamber center, a series of optics combined the beams from both lasers to produce a single beam that spanned the chamber center. A quarter-wave plate converted the cross-polarized laser beams to circular polarization for balanced particle scattering between pulses, and a concave cylindrical lens and square aperture sized the imaging volume as a $50\text{mm} \text{ by } 30\text{mm} \text{ by } 5$mm box. The illuminated particles were simultaneously captured by four identical high-speed cameras in frame-straddling mode (Phantom Veo 640L, 2560 \text{by} 1600 pixels, 200mm macro lenses, $f$/27) positioned at different perspectives to triangulate the 3D positions of particles. The cameras were positioned $20^\circ$ from the normal direction of the laser sheet and oriented in a cross-configuration (Fig. \ref{fig:01}). The effective pixel scale was 21$\mu$m. With a working distance of $0.7$m for the 0.5m radius flow facility with 0.2m lens, it was critical to isolate for vibration, since miniscule incidental deviations of individual camera angles would lead to pixel-level deviation of pixel positions from the calibration.\\

\textbf{Vibration mitigation} - The four cameras were rigidly mounted on a passive vibration-isolating table, such that vibrations from external sources such as the turbulence chamber fan motors and building vibration were damped. The table has a natural frequency of 3Hz, so any undamped swaying motion of the table occurs over a timescale much larger than the $200\mu s$ -duration recordings. Sways between recordings did not affect the statistics of $r$ and $w_r$, since the sway was identical among the cameras, leading to translation of coordinate origin that is independent of $g(r)$ and $w_r$. To minimize breezes that might incidentally move lenses, the lasers were isolated from the cameras and lab ventilation was diverted during data collection.

\subsection{Implementation of Shake-the-Box Particle Tracking}
We implemented our small-$r$ measurement strategy using the multi-pulse STB tracking algorithm \citep{sellappan2020lagrangian,RN19} based on Shake-the-Box \citep{RN21} implemented in DaVis 10.1 by LaVision GmbH (G\"{o}ttingen, Germany), followed by an in-house particle-pair mismatch rejection code described in Section 4. The STB particle tracking algorithm works by triangulating particle positions using an array of cameras and includes a unique approach to refine the particle position using the particle images, which makes it advantageous for small-$r$ measurements. Important to our high-resolution measurements, the distance by which a particle is ``shaken" in each iteration (0.1 pixels in our experiments) plays into the resolution limit of STB, since it acts as a precision limit of the particle position. To minimize calibration error from small drifting of the camera and lens mounts, we performed the volume self-calibration with the images used to calculate RDF and RV. The final average disparity between self-calibration iterations was $<$ 0.1 voxel ($\approx 2\mu$m), as recommended by \cite{RN22}.

\par The timing scheme is shown in Fig. \ref{fig:01}. We chose $\Delta t_1=\Delta t_3=1.6\Delta t_2$ based on the suggestion of $\Delta t_2 < \Delta t_{1,3}$ as a suitable choice for the recording of multi-exposed images for STB \citep{RN19,sellappan2020lagrangian}, and based on minimizing $\Delta t_2$ to reduce uncertainty at small $r$ due to interpolation error (see Section \ref{sect:07}). To allow for a maximum of 10 pixels of particle displacement between frames for a high dynamic range as recommended by LaVision, $\Delta t_1$ and $\Delta t_3$ were chosen to be $70\mu s$ based on the RMS velocity of the flow measured as $1.2$m/s \citep{RN20}, such that $\Delta t_2$ was chosen as $44 \mu s$. To achieve statistical independence between recorded realizations, the repetition frequency of the four pulses was set at the lowest camera frame rate (12 Hz) such that the time between realizations was $83$ms, as compared to the large eddy turnover time of $150$ms \citep{RN20}. 

\par In this paragraph, we list detailed values of our 4P-STB inputs. The threshold for 2D particle detection was 70 counts (out of 4096). The maximum allowable triangulation error $\epsilon$ was $1.5$ voxel (voxel size $\approx 21\mu m$). We used four iterations of the inner and outer shaking loops, with a shake size of 0.1 voxel. Particles were removed if found to have $r<0.7$ voxel, as this condition was physically impossible for the particles in our experiments. Only a single iteration of IPR was performed. To calculate an optical transfer function (OTF) for use in STB, the flow was divided into 50 equally-sized sub-volumes (5 by 5 by 2). For each sub-volume and camera angle permutation (50 subvolumes by 4 cameras), a single OTF was generated to represent the particles in each subvolume, as seen by the camera. The original recorded particle images were then used to fit a an OTF by finding the optimal values of weighting functions $x_0$, $y_0$, $a$, $b$, and $c$ as described in \cite{schanz2012non}. The OTF is then used in STB as detailed in \cite{RN19,sellappan2020lagrangian}.

\par Using the above described 4P-STB technique, we recorded particle tracks in $\mbox{15 465}$ realizations of the isotropic turbulent flow to ensure convergence of RDF and RV. The average and standard deviation of the number of analyzed particles in each realization were 434 and 148, respectively. Each four-pulse track provides one instantaneous particle position and velocity. The particle positions and velocities were averaged over all realizations to calculate RV, followed by RDF.

\section{Relative Velocity PDFs Calculation and Results}
\label{sect:04}
\textbf{Radial relative velocity calculation} - For particle $A$ and $B$, their radial relative velocity is defined as $w_r=(\mathbf{v_A}-\mathbf{v_B})\mathbf{\cdot}(\mathbf{r}/|\mathbf{r}|)$, where $\mathbf{v_i}$ is the velocity vector of particle $i$, and $\mathbf{r}=\mathbf{x_A}-\mathbf{x_B}$, where $\mathbf{x_i}$ is the position vector of particle $i$. For each realization of the turbulence, $w_r$ of every particle pair in the flow was calculated. These $w_r$ values were then binned by $r$ for $2.07<r/a<650$ (equivalent to $0.24<r/\eta<81.1$) into 91 bins. The bins were logarithmically spaced and chosen to resolve the tails of PDFs at the smallest separations. For each bin of $r$, we calculated the PDF of $w_r$, $p(w_r(r))$. Figure 2 shows five representative PDFs at $r/\eta = (0.24,0.44,0.92,1.68,10.2,30.0)$, which correspond to $r/a = (2.07,3.78,7.91,14.46,88.4,259)$.

\begin{figure}
	\centering
	\includegraphics[scale=0.9]{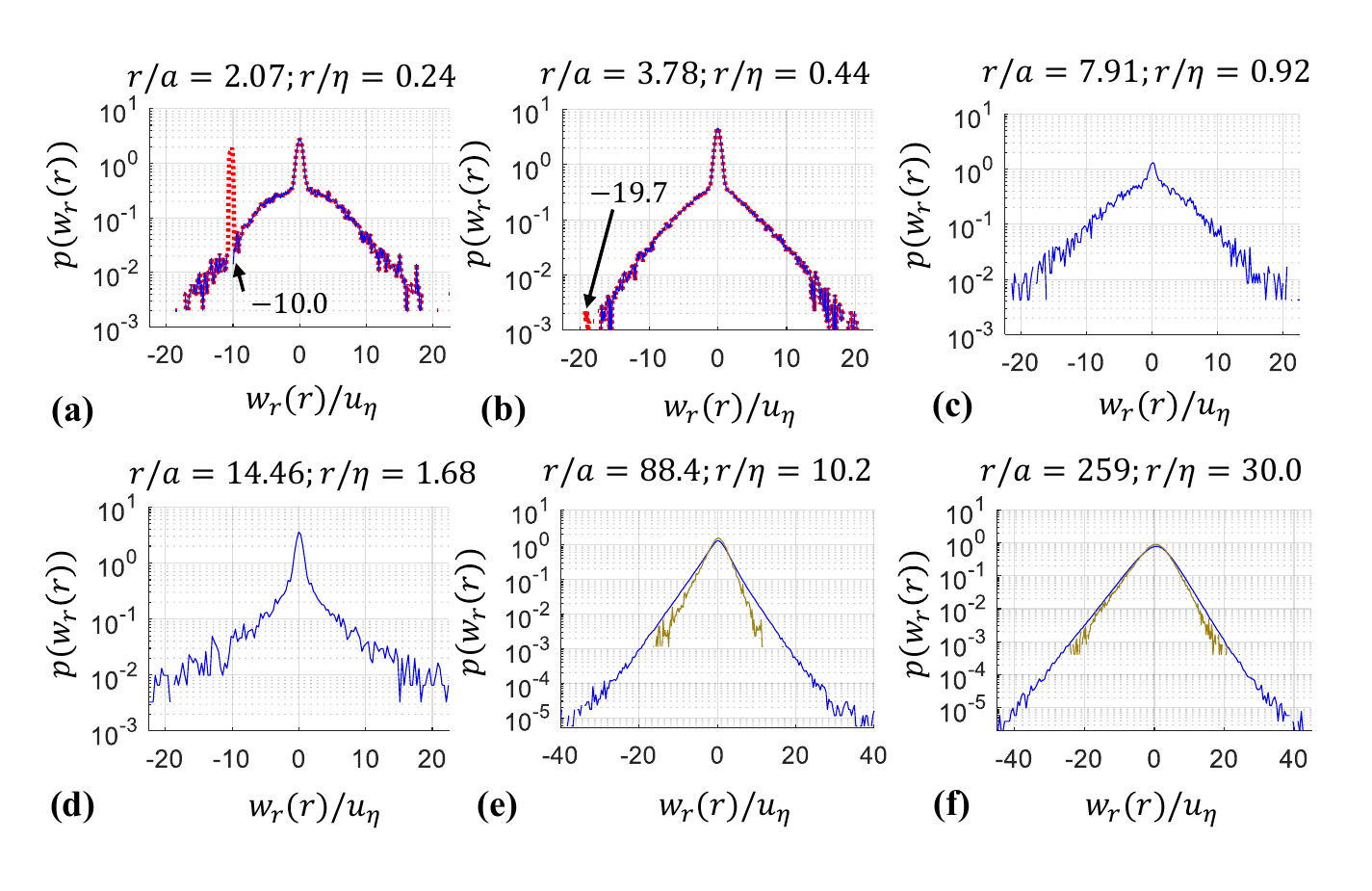}
	\caption{Probability density functions of particle-pair relative velocity at six separations. (a) $r/a=2.07$ (near contact). The prominent dashed-line peak was due to particle mismatches $w_{mis}/u_\eta = -10.0$ m/s Correcting for mismatch results in the blue curve. (b) r/a=3.78. The peak due to mismatches diminished. (c) $r/a=7.91$. (d) $r/a=14.46$. (e) $r/\eta=10.2$ compared with \cite{RN16}. (f) $r/\eta=30.0$ Compared with \cite{RN16}, with more prominent negative skewness. \sdasha, Before mismatch removal (current study); \sdashb, after mismatch removal (current study); \sdashc \cite{RN16}.}
	\label{fig:02}
\end{figure}

\textbf{Removal of particle mismatch} - The particle number density used in this study is small (0.002 particles per pixel, ppp), such that in general, particle tracking error is not expected to be prevalent. For small-number-density cases, tracking ambiguity may still occur when pairs of particles are extremely near to one another. The result of this tracking ambiguity is that, although rare, the tracking algorithm may swap the identity of the tracked particle with its neighbor, leading to erroneously crossed particle tracks. We term this track-swapping phenomenon as ``mismatch". When tracks erroneously cross, the apparent particle separation at the crossing is extremely small. If this swap occurs between pulses 2 and 3, this will lead to an erroneous, near-contact separation at the track midpoint. Because of the swap, there will also be a false inward RV from the false ``relative velocity" from the pairs switching places. When this occurs, the relative velocity will appear as $w_{mis}=(-2r_{mis})/\Delta t_2$. This expression comes directly from the tracking algorithm switching the particle positions: a false inward displacement of the particles $(-2r_{mis})$ has been manufactured over the track interpolation time $(\Delta t_2)$ by the tracking algorithm. 

We use $w_{mis}$ to identify and remove mismatched tracks. The first pass of $p(w_r(r))$ calculation is shown in Fig. \ref{fig:02}a and b as the red dashed curves. The sharp spike at $-1.34$m/s for $r/a=2.07$ in Fig. \ref{fig:02}a was exactly $w_{mis}$. After removing particles with $w_r=w_{mis}$ from each PDF, we obtained the corrected RV PDFs for all the conditions, exemplified by the blue curves in Fig. \ref{fig:02}. For $r/a \gtrapprox 3.78,\:w_{mis}$ was beyond the maximum measurable $w_r$ based on the dynamic range of the velocimetry system and therefore its removal was inconsequential. 

\textbf{Relative velocity PDF result discussion} - As exemplified by Fig. \ref{fig:02}a and 2b, all the RV PDFs for $r/\eta\lessapprox 0.5$ exhibit a prominent narrow core abruptly transitioning to broad tails. This suggests that there could be two additive mechanisms driving the particle relative velocity at small separations. In contrast, as demonstrated in Fig. \ref{fig:02}e and 2f ($r/\eta=10.2$ and $r/\eta=30.0$), the PDFs at very large separations do not exhibit a core-and-tail structure, though there is a slight upturn visible at the most extreme values of $w_r(r)$. Starting from near-contact, when $r/\eta$ increases to $\approx 0.5$, the core remains qualitatively the same, but the curvature of the tails decreases (see Fig. \ref{fig:02}a and 2b). As $r/\eta$ further increases to approximately 1, the core becomes obscured by the rise of the tails (see Fig. \ref{fig:02}c). As $r/\eta$ continues to increase beyond 1, the tails drop lower, revealing a structurally different core with smooth transitions to the tails. With further increase of $r/\eta$, the tails diminish, leaving the linear-in-the-log-scale core to widen (see Fig. \ref{fig:02}e and 2f).

\par To compare our RV results against the literature, in Fig. \ref{fig:02}e and 2f we co-plot our results with the RV PDF from the experimental measurement by \cite{RN16} under the same flow and particle conditions and thus the same $St$ and $\Rey_\lambda$ as in the current study. However, \cite{RN16} used a different, 2D particle tracking technique. Both PDFs show a linear core shape in the log-scale, but that by \cite{RN16} was narrower. This is likely due to their 2D technique (as opposed to our 3D technique), which led to underprediction by $\sqrt{2/3}$.

In Fig. \ref{fig:02}e and 2f, we observe the RV PDFs to be slightly negatively skewed. This is expected as a result of vortex stretching in turbulence \citep{tavoularis1978velocity}. In smaller separations (Fig \ref{fig:02}a-d) the PDFs become symmetric. Compared to the RV PDFs of \cite{RN15} who did observe skewness in their RV PDFs at $r/\eta \approx 1$, the tails of our RV PDFs are much higher. This means that we observed larger RV values more frequently than they did in their experiments, which may have overshadowed the less-frequent negative skewness effects caused by vortex stretching. It should be noted that our experiments were under very different conditions (e.g. larger $a$, smaller density $\rho$, smaller $\eta$, solid particles) compared to those of \cite{RN15}. These different conditions could have caused PPI to occur at larger $r/\eta$ in our experiments than in \cite{RN15}.

\begin{figure}
	\centering
	\includegraphics[scale=0.78]{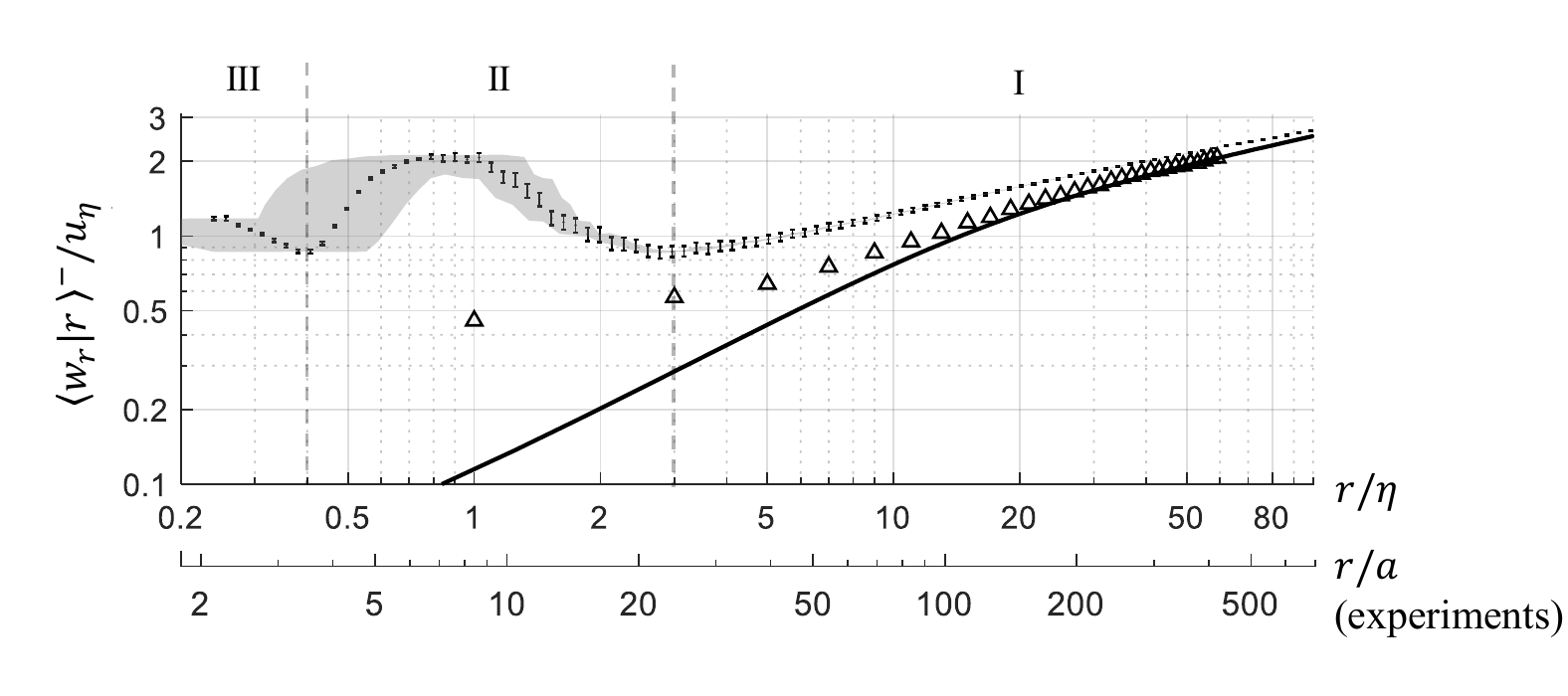}
	\caption{$\langle w_r|r\rangle ^-$ normalized by $u_\eta$, compared with \cite{RN16} and DNS of \cite{RN4}. The shaded region represents the uncertainty due to interpolation, interpreted as the range of $r$ which may contribute to the measurement. The vertical error bars are calculated as in Section 7, uncertainty by ensemble forecast. From the right, Regions I, II, and III are characterized by the monotonic decrease due to turbulence, a plateau, and an increase towards contact, respectively. 
		\sstar Current Study, \striangle Experiment by \cite{RN16}, \sdash DNS by \cite{RN4}.}
	\label{fig:03}
\end{figure}

\textbf{Inward RV result} - For the collision kernel in Eq. \ref{eqn:01}, we calculated the first moment of negative velocities $\langle w_r|r\rangle^-=\int_{-\infty}^{0}{w_r p(w_r|r)dw_r}$ and plotted it against $r/\eta$ and $r/a$ in Fig. \ref{fig:03} (vertical bars), along with experimental results by \cite{RN16} (triangles) and DNS results by \cite{RN4} (solid line). The vertical error bars in the new experimental results were calculated as described in Section 7: Uncertainty by ensemble forecast. As $r/\eta$ decreases from 80 to 3 (denoted as \textbf{Region I}), our measured $\langle w_r|r\rangle ^-$ decreases monotonically, consistent with previous results. However, at $r/\eta=3 \text{ } (r/a=25)$, the newly measured $\langle w_r|r\rangle ^-$ turns upward with decreasing $r$. When $r/\eta\sim 1$ it plateaus. After $r/\eta\sim 0.6$ it decreases again, reaching a minimum at $r/\eta=0.4$. We denote this region, $0.4<r/\eta<3$ (which corresponds to $3.3<r/a<25.9$) as \textbf{Region II}. When $r$ decreases further towards contact, $\langle w_r|r\rangle ^-$ increases again, reaching approximately $1.2u_\eta$ at $r/a=2.07$. This is denoted as \textbf{Region III}. The shaded regions around our measurement data represent horizontal uncertainties arising from track interpolation (detailed in Section \ref{sect:07}). Note that the DNS by \cite{RN4} assumed one-way coupling (no PPI) for monodisperse inertial particles at $St=0.7$ and $\Rey_\lambda=398$.

\textbf{Inward RV result discussion} - Figure 3 shows that at very large $r$ in Region I, all results overlap. At these scales, turbulence alone drives the particle relative motion. As such, the DNS of non-interacting particles match the experiments. As $r$ decreases to $r/\eta=20$ in Region I, both experimental studies show higher $\langle w_r|r\rangle ^-$ than the DNS. This could be due to the presence of weak PPI, which is not accounted for in the DNS. Between the two experiments, the results of this study are higher than those of \cite{RN16}, due to difference between the 3D and 2D measurements. \cite{RN16} speculated particle polydispersity in their experiments as the cause for their elevated $\langle w_r|r\rangle ^-$ compared to DNS. However, we believe that polydispersity effects are not dominant until $r$ decreases to Region III. 

Starting in Region II ($r/a\approx 25$), the inward relative velocity begins to increase,  indicating a decorrelation of the particle relative motion. Qualitatively, the increase of $\langle w_r|r\rangle ^-$ is reminiscent of inward drift by hydrodynamic interactions between inertia-free particles \citep{RN25}. However, since our particles have appreciable inertia, the interactions between them may be more complicated than the HI predicted for inertia-free particles by \cite{RN25}. The measured inward RV then peaks at around $r/a\approx 10$ and decreases thereafter. \cite{RN25} explained that lubrication suppresses the relative velocity between particles. Their theory predicted a peak at $r/a\approx 2.08$; however, our data peaks at $r/a\approx 10$. If the peak and downturn we observed was from lubrication, this would mean that lubrication is acting across longer distances than the prediction by \cite{RN25}.

In Region III, $\langle w_r|r\rangle ^-$ is enhanced again, which we believe is an effect of polydispersity on the particle motion in the flow. Particles of different sizes will respond to the flow differently, thus enhancing their relative velocities. Although we aimed to produce monodisperse particles by sieving (as described in \citep{RN17}), the sieved particles have a narrow but finite size distribution. When $r$ decreases to the scales of multiple radii, the minute difference in particle size will lead to the enhancement of relative velocity. Relative velocity enhancement due to dispersion of particle size has been previously observed in simulations of bidisperse, noninteracting particles \citep{zhou2001modelling}.

\section{Radial Distribution Function Calculation and Results}
\label{sect:05}

\textbf{RDF calculation} - The RDF measures the degree of particle clustering in the flow. It compares the expected number of satellite particles at a distance $r$ from each primary particle to the number of expected satellite particles in a uniform spatial distribution. It can be calculated by binning the particle pairs according to their separation distance, and then calculating \citep{RN14} $g(r_i)=\frac{N_i/(\Delta V_i)}{N/V}$, where $N_i$ is the number of particle pairs separated by a distance of $r_i \pm \Delta r/2$ and $\Delta r$ is the width between the bins. $\Delta V_i $ is the volume of a spherical shell of radius $r_i$ and thickness $\Delta r$, $N$ is the total number of particle pairs, and $V$ is the overall volume of the flow. Using this approach, we calculated the RDF for each of the $15\text{ }465$ realizations, then took the ensemble average of all the RDFs as the result.

\textbf{RDF boundary treatment} - When RDF is calculated in a finite sample volume, it is paramount to properly treat the boundary to obtain accurate estimation of $g(r)$ at large $r$. Without it, those primary particles near the edge would not have satellite particles to pair with in space outside the imaging volume. This would lead to an underestimation of $N_i$ that affects more particles as $r$ increases, leading to diminishment in $g(r)$ which intensifies as $r$ increases.
\par A recent study by \cite{RN23} discussed the boundary treatment for RDF in depth, outlining two methods suitable for RDF experiments: the guard area approach, which allows the user to define the volume within a distance $\delta x$ from the boundary edges wherein particles may only be considered as satellites for pairing, and their new effective volume approach, which accounts for the edge-effects of primary particles near the volume boundary and does not exclude these particles. The former is computationally inexpensive but loses data, while the latter retains data for statistical convergence but is computationally expensive when used at high resolution. For our boundary treatment in RDF calculations, we combined these two strategies: we used a $\delta x=0.5$mm guard area for $0<r\leq 0.5$mm and the effective volume approach for $r>0.5$mm.

\begin{figure}
	\centering
	\includegraphics[scale=0.7]{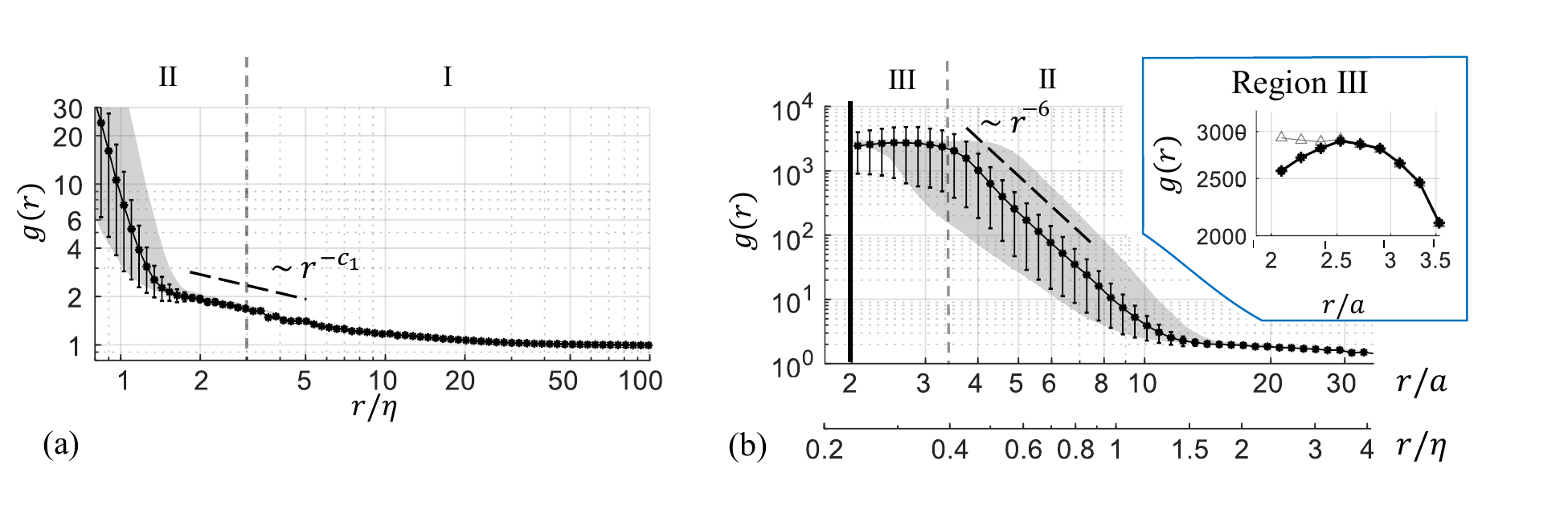}
	\caption{Measured RDF of particles in isotropic turbulence ($St=0.74$, $\Rey_\lambda=324$). The shaded area represents the uncertainty in $r$ from interpolation uncertainty, i.e. the range of $r$ which may contribute to the measurement. The vertical error bars are uncertainty by ensemble forecast calculated in Section 7. (a) $g(r)$ against $r/\eta$ in the range $0.8\leq r/\eta\leq 100$. (b) $g(r)$ against $r/a$ in the range $2.07\leq r/a\leq 34$, where the solid vertical line represents contact. Regions I, II, and III are separated by dashed vertical lines. The insert shows the effect of mismatch removal in Region III, near contact. \striangledash $g(r)$ before mismatch removal; \sstardash $g(r)$ after mismatch removal. All axes are log-scaled.}
	\label{fig:04}
\end{figure}

\textbf{RDF result} - Using the particle position data from the particle tracks, we calculated the radial distribution function using boundary treatment. The resulting RDF is plotted in Fig. \ref{fig:04}. To visualize the scaling from large $r$ down to $r\approx \eta$, in Fig. \ref{fig:04}a, $g(r)$ is plotted against $r/\eta$. Furthermore, to examine $g(r)$ at small separations down to contact, in Fig \ref{fig:04}b, we replot $g(r)$ against $r/a$ for $2.07<r/a<35$, which corresponds with $0.23\leq r/\eta\leq 0.4$. The shaded region on $g(r)$ represents the error bounds of $r$, which reflect the interpolation effect on the measurements, as detailed in Section \ref{sect:07}. The vertical error bars are uncertainty by ensemble forecast calculated in Section \ref{sect:07}.

\par The entire regime of $r$ can be divided into Regions I, II, and III consistent with the RV plot (Fig. \ref{fig:03}). In large scales (Fig. \ref{fig:04}a), $g(r)\rightarrow 1$ as $r/\eta \rightarrow \infty$, as expected in isotropic flows. As $r$ decreases across Region I and a part of Region II, the RDF increases by a well-known power law scaling $g(r)\propto r^{-c_1}$, evidently due to the preferential concentration effect \citep{RN24}. As $r$ further decreases to $r/a\approx 12$, which is $r/\eta \approx 1.5$, the RDF starts to exhibit a surprising explosive increase. As $r$ goes below $r/a \approx 3.5$ (Region III), the RDF plateaus.

\textbf{Effects of mismatch removal} - Particle mismatch over the track midpoint not only caused a spurious spike in RV PDF as shown earlier, but also artificially increased RDF at near-contact $r$. The inset of Fig. \ref{fig:04}b shows that particle mismatch occurred at $r/a<2.75$, and mismatch removal led to a 20\% correction (reduction) in the near-contact RDF. For $r/a \gtrapprox 2.75$, mismatch removal made no difference on the RDF estimates.

\textbf{RDF result discussion} - A power law fit of the $r^{-c_1}$ regime observed in the RDF yielded $c_1=0.39$, which is smaller than $c_1=0.69$ reported by DNS from \cite{RN4} for monodisperse particles at $St=0.7$. Our measured
 $c_1$ value is reasonable, since the particles in the experiments are polydisperse, and polydispersity diminishes the value of $c_1$ for the overall particle sample based on the least clustered particle population \citep{saw2012spatial1,saw2012spatial2}. In our experiment, the least clustered population is comprised of the smallest $St$ particles in the sample. To properly compare the experimental $c_1$ against DNS, simulations would need to use the same particle radius distribution as in the experiment.

\par Theoretical models of far-field particle-particle hydrodynamic interactions of inertia-free particles show that the pair probability $\rho(r)$, which is proportional to $g(r)$, scales with $r^{-6}$ \citep{RN25}. This scaling arises by solving their Eq. (28) using the far-field forms of their functions. In Fig. \ref{fig:04}b we observe a clear $r^{-6}$ scaling in $g(r)$. This suggests that HI may be dominating RDF in Region II in our experiments, even though \cite{RN25} predicted $r^{-6}$ scaling for inertia-free particles. Incidentally, \cite{RN18} reported a strong upturn in $g(r)$ near $r/a\approx 10$ similar to our experiment, but did not report any $r^{-6}$ scaling. Instead, they used theoretical analyses to infer that the $r^{-6}$ scaling regime would have occurred at a smaller $r$ than their experiment could resolve. We hold the opinion that their data in these small separations could well have embedded  $r^{-6}$ scaling, except that it was obscured by their experimental noise evidenced by the large scatter of their data. 

\par At the start of Region III ($r/a\lessapprox 3.3$), $g(r)$ starts to plateau. This is likely due to particle polydispersity discussed above for inward RV in the same region. DNS of noninteracting particles have shown that polydispersity diminishes the turbulence enhancement of $g(r)$, leading to $g(r)$ plateauing at small $r$ \citep{saw2012spatial1,saw2012spatial2,RN26}. Similarly, we suspect that polydispersity also diminishes the PPI enhancement of $g(r)$, albeit at even smaller $r$. In both cases, the effect of polydispersity is to decorrelate particle responses to the local flow. For turbulence, this decorrelation arises due to the varying levels of inertia of the particles (Saw et al. 2012a). For PPI such as HI, we suspect that this decorrelation arises due to the varying sizes of the particles.

\section{Enhancement of RV, RDF, and Collision Kernel}
\label{sect:06}
\textbf{RV enhancement} - Figure 3 shows that as $r$ decreased, our experimentally-measured inward RV $\langle w_r(r)|r\rangle ^-$ turned upward at the border between Regions I and II, insead of continuing the monotonic decrease predicted by the DNS \citep{RN4}. Other DNS studies also predicted similar monotonic decreases \citep{wang2008turbulent,rosa2013kinematic}, even when including a quasi-steady Stokes flow model for HI (termed aerodynamic interactions in their reports). \cite{wang2008turbulent} reported that the addition of HI marginally weakened the inward RV and did not change its monotonic decrease as $r$ decreased. Our experimental measurement of inward RV in Region II shows a more complex behaviour. Since our experiment does not use simplifying assumptions and captures the full physics over the range that the experiment can resolve, we conjecture that previous simulations did not fully account for particle interactions.

\textbf{RDF enhancement} - At $r/a$=12 (or $ r/\eta=1.5$), the RDF value is still comparable to previous experiments and DNS results by \cite{RN14}, but the immediately following $g(r)$ enhancement by PPI that scales as $r^{-6}$ brings RDF all the way to a staggering 2 000 at $r/a=3.5$. The near-contact $g(r)$ is thus $O(10^3)$, compared to extrapolation from the $r^{-c_1}$ scaling from PTI alone, which was $O(10)$. Prior DNS without HI \citep{wang2000statistical,RN4} and with a model for HI \citep{wang2008turbulent,rosa2013kinematic} also predicted a power law scaling exponent that leads to a near-contact $g(r)$ of $O(10)$. This suggests that our experimental data may contain physics not captured by prior models.

\textbf{Collision kernel} - From our near-contact RDF and RV data, a collision kernel that retains PPI (hydrodynamic interactions included, in absence of the Coulomb force) can be calculated. To compare with DNS, we calculated the nondimensional collision kernel $\hat{k}(2a) = k(2a)/(2a)^2u_\eta$ following \cite{RN4}. From the smallest measured separation $r/a=2.07$, we extrapolate RDF and RV down to $r/a =2.00$, obtaining $g(r)|_{r=2a}=2500$ and $\langle w_r|r=2a\rangle ^-=1.2$m/s at contact. This yields $\hat{k}(2a)=2.9\times 10^5$ for $St\approx 0.75$, $a\approx 14\mu$m, $\Rey_\lambda=324$. Since RDF is enhanced far more than RV, it is evident that PPI enhances collision rate mostly through increasing clustering.

While our measurement resolution allowed for probing the statistics at very small separations (down to $r/a=2.07$), the measured $r$ is subject to interpolation uncertainty in particle tracking (Section \ref{sect:07}). This uncertainty in $r$ will affect the statistics of $g(r)$ and $\langle w_r|r\rangle^-$, which depend on $r$. Consequently, effects of physics that drive particle RV or RDF will be averaged over the $r$ uncertainty, which could be wider than the relevant $r$ of the physics itself. For example, the breakdown of the fluid continuum assumption occurs at separations on the order of the mean free path in air, which is much less than $r/a=2.07$. The masking of this effect could lead to error in the extrapolation to contact for calculation of the collision kernel. Indeed, to to accurately estimate the collision kernel at contact, all near-contact physics need to be accounted for, which is beyond the current experimental capabilities. However, our experiments have already pushed the envelope for future modelling by providing collision statistics at much closer-to-contact separations, allowing collision kernel estimations that are more credible than extrapolation from the PTI-dominated regime (Region I).

Our calculated collision kernel is $4$ to $6$ orders of magnitude higher than DNS predictions by \citet{RN2} and \citet{RN4} under the one-way coupling assumption and neglecting PPI, which are $\hat{k}(2a)=0.1-10$ for $St\approx 0.1-1.0$ and $\Rey_\lambda=88-597$. This astounding collision enhancement provides experimental evidence that PPI drastically increase particle collision rates.
Evidently, prior models of HI implemented in simulations did not fully capture the extent of the enhancement of collision rates as observed in this experiment. As mentioned above, simulations with prior HI models (e.g. \cite{wang2008turbulent}) obtained results that were functionally similar to the DNS results without any PPI, with only slight magnitude differences. Thus, the models of PPI used in the past for simulating collision statistics in turbulent flows may not fully reflect the true nature of particle interactions at near-contact separations.

\section{Measurement Uncertainty}
\label{sect:07}

\begin{figure}
	\centering
	\includegraphics[scale=0.7]{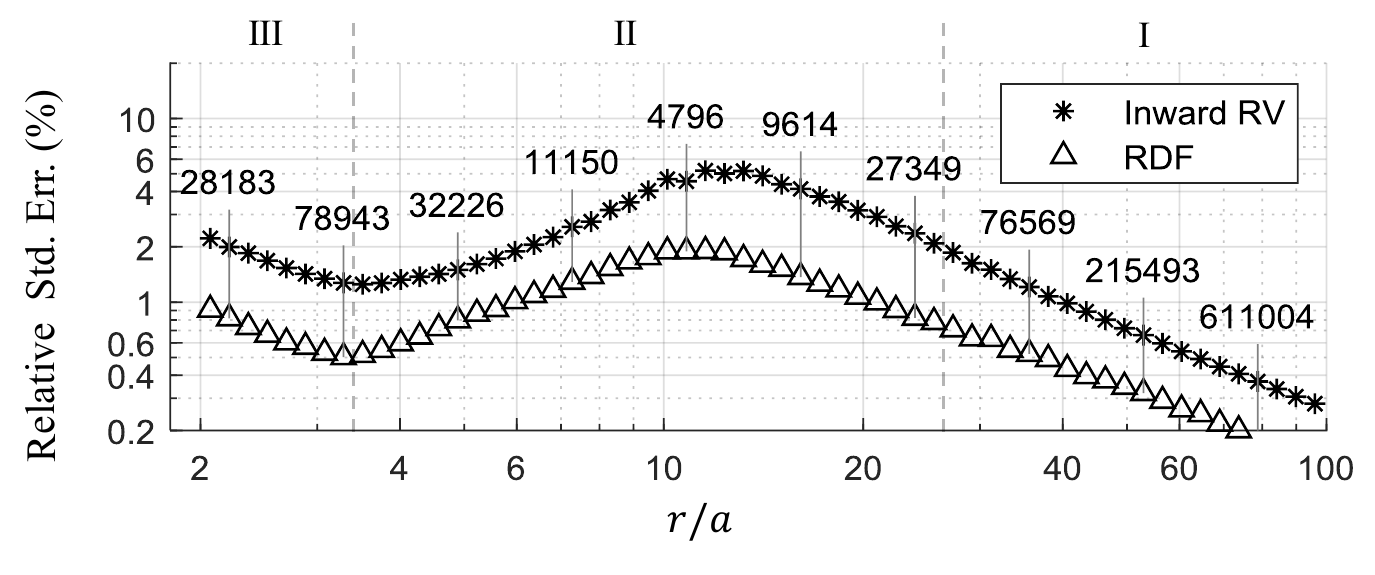}
	\caption{Figure 5. Relative standard error of the mean for Inward RV (\sstardash) and RDF (\striangledash) as a function of separation. The dashed vertical lines separate Regions I through III. The numbers above the vertical-line-marked symbols correspond to the total number of recorded particle pairs for the marked bin. The number of counts per bin increases as $r/a$ increases beyond 100. The total number of particle pairs for the entire experiment was $1.19\times 10^9$.}
	\label{fig:05}
\end{figure}

\textbf{Sample size and statistical convergence} - To ensure that the experimental results were statistically significant, we aimed to acquire sufficient experimental data to converge the RV and RDF statistics with minimal standard error. The data was taken over $15\text{ }465$ realizations, with on average 434 particles per frame. The sample size of particle pairs in a given bin of separation for calculation of the RV and RDF ranged from $O(10^3)$ to $O(10^6)$. In Fig. \ref{fig:05} we plot the relative uncertainty based on the standard error of the mean for both inward RV and RDF along with the sample size at select $r$ bins. We find that for both statistics, the standard error always remains below 5\%. The highest standard error and occurs not at $r$ near contact, but in Region II, where the sample size is also the lowest. This corresponds to the beginning of the $r^{-6}$ upturn in RDF, and the regime where inward RV increases for decreasing $r$. Due to the clustering, there are fewer particle pairs at these intermediate separations.

\textbf{Sample size of the removed mismatches} - To ensure that the complete removal of the mismatches (described in Section \ref{sect:04}) did not affect the RV and RDF statistics, we compared the number of mismatches to the total sample size. There were a total of 8277 mismatched particle pairs out of $1.19\times 10^9$ total particle pair samples. 97\% of these were found across the first 5 bins. The percent of mismatches dropped quickly from 15\% (first bin, $r/a=2.07$) to 0.3\% (fifth bin, $r/a=2.70$). The removal of these mismatches did not affect the RV and RDF in these bins because the true separations of the mismatches were larger than $r_{mis}$ (See \ref{sect:04}, Removal of particle mismatch). The bins belonging to the true separations of these particles have orders of magnitude more data than even the total number of mismatches, and thus removal of these mismatches are inconsequential.

\textbf{Interpolation uncertainty} - Our track interpolation technique allowed us to obtain $w_r(r)$ and $g(r)$ at much smaller $r$ than previously possible. However, the accuracy of $r$ and $w(r)$ at the track midpoint is still limited by the spatiotemporal resolution of the experimental setup. Despite the interpolation time $\Delta t_2$ being small, it is still finite. Since there is a relative velocity between particles, the instantaneous value of $r$ varies over the interpolation time $\Delta t_2$. This means there is uncertainty in the true $r$ at the time of track midpoint due to the interpolation. In other words, if particle pairs have fluctuations in their relative position over the track that occur over timescales smaller than $\Delta t_2$, the particle track recovered by 4P-STB will not reflect these fluctuations. 

To quantify this interpolation uncertainty, we calculated the RMS radial distance travelled by particles between the second and third pulse of the four-pulse track as $\delta r_{in}=\Delta t_2\sqrt{\langle w_r^2(r)\rangle }$, where $\langle w_r^2(r)\rangle $ is the variance of the particle-pair radial relative velocity PDF. This affords an estimate of the range of $r$ values, with potentially different physics, that may contribute to the data used to calculate RDF and RV at a given $r$ bin. When $\delta r_{in}$ becomes comparable with $r$, the interpolation uncertainty must be considered to interpret the results. 

\textbf{Confidence interval based on interpolation uncertainty} - To account for the effect of interpolation over $r$, a confidence interval $r \pm \delta r_{in}$ is added as the shaded regions (consisting of horizontal bars) in $g(r)$ and $\langle w_r|r\rangle ^-$ in Figs. 3 and 4. Clearly, interpolation uncertainties are negligible at large $r$, but as $r$ decreases to the order of $\eta$, the confidence intervals start to widen. 

A question then arises as to whether the upward and downward trends of these curves are real. Note that in Fig. \ref{fig:03}, even at contact, particle pairs cannot be misconstrued as pairs separated by $r/a\approx 4$ with the same value of $\langle w_r|r\rangle ^-$, since the confidence intervals at $r/a \approx 4$ do not reach as far as $r/a=2$ (contact), and vice versa. Likewise, in Fig. \ref{fig:04}, the $r^{-6}$ scaling of $g(r)$ and the $r$ values for the plateau are significant. Hence, we believe that all the observable trends in $g(r)$ and $\langle w_r|r\rangle ^-$ are real. However, when the particles are nearly in contact, the relevant timescale of particle interaction $\tau_x$ should diminish, and the physics dominating timescales $\tau_x<\Delta t_2$ is not captured. This may include lubrication forces, which dampen relative velocities extremely near to contact, i.e. when $r/a\approx 2.08$ for inertia-free particles \citep{RN25}. This is a current limitation of our technique.

\textbf{Particle position uncertainty} - The recorded particle positions from the tracks themselves have uncertainties, which affect the precision of $r$ and thus $w_r(r)$ and $g(r)$. Owing to our careful measures to acquire high-quality tracks through vibration isolation and volume self-calibration on the images used to collect data, we expect that our position uncertainty is on par with 0.15 pixels \citep{RN19}, which, based on a camera pixel size of $21\mu$m, translates into a position uncertainty $\delta x =3.2\mu$m and an $r$ uncertainty of $\delta r=\sqrt{2}(0.15)(21\mu \text{m})\approx 4.5 \mu$m estimated via propagation \citep{RN27}. These uncertainties, only a fraction of the particle radius, are overshadowed by the interpolation uncertainty.

\textbf{Uncertainty by ensemble forecast} - The 4P-STB particle tracking algorithm has many user-defined parameters. Among these, the allowable triangulation error $\epsilon$ is considered the most important and consequential parameter affecting the output \citep{RN19}. We used $\epsilon=1.5$ voxel (1 voxel $\approx 21\mu m$), since this value produced the most total tracks when tested against nearby values. To test the sensitivity of RV and RDF to variation in $\epsilon$ and acquire vertical error bars, we varied $\epsilon$ by $\pm$ 10\% and calculated inward RV and RDF at $\epsilon=1.35$ and 1.65 voxels. For each separation, we then took twice the standard deviation of the results ($\epsilon=$ 1.35, 1.5, and 1.65 voxels) as the vertical error bar for the inward RV and RDF, shown in Figs. \ref{fig:03} and \ref{fig:04}. This process is akin to ``ensemble forecasting" in weather prediction, where multiple different forecasts are produced with different input conditions to estimate the range of potential weather outcomes.

The resulting error bars show that at near-contact the RV was not strongly affected by the triangulation error. At this separation, the vertical error bar from the ensemble forecast uncertainty was only 2.1\%, similar to its standard error (2.3\%). On the other hand, for the RDF uncertainty at near-contact, there is potential for 60\% variation in the experimental result of RDF, even though the standard error is 1\%. However, this variation does not change the order of magnitude of the predicted RDF.

\section{Conclusions}
\label{sect:08}
We report the first-ever detailed, simultaneous measurement of RDF and RV at much small $r$ down to near-contact for experimental estimation of the collision kernel. Based on a 4P-STB particle tracking technique, our novel track-midpoint particle positioning approach aided by a mismatch rejection algorithm has allowed acquisition of particle positions at much smaller $r$ than previously possible, leading to observations of dramatic enhancements of inward RV and RDF. The data reveal 3 distinct regions of particle separation distance: PTI-dominated \textbf{Region I} (down to $r/\eta=3$): PPI-dominated \textbf{Region II} ($0.4<r/\eta<3$), containing three inversions in $\langle w_r|r\rangle ^-$ and $r^{-6}$ scaling in $g(r)$; and PPI-dominated \textbf{Region III}, where $\langle w_r|r\rangle ^-$ increases and $g(r)$ plateaus due to polydispersity. The resulting non-dimensional collision kernel is 4-6 orders of magnitude higher than predictions by DNS, which do not model PPI. We hope that the new experimental data from this study will stimulate more investigations of near-contact physics and thereby help improve modeling of particle collision statistics accounting for PPI.

\section*{Acknowledgements}
We thank the National Science Foundation for support through the Major Research Instrumentation program (Award 1828544, Program Manager Harsha Chelliah). We are grateful to Andrew D. Bragg for stimulating discussions and critique, Danielle Johnson for editorial assistance, and Lance Collins and Raymond Shaw for helpful discussions. We thank LaVision for their continuous technical support.

\section*{Declaration of Interests}
The authors report no conflict of interest.

\bibliography{jfm}

\begin{thebibliography}{42}
\providecommand{\natexlab}[1]{#1}
\providecommand{\url}[1]{\texttt{#1}}
\expandafter\ifx\csname urlstyle\endcsname\relax
  \providecommand{\doi}[1]{doi: #1}\else
  \providecommand{\doi}{doi: \begingroup \urlstyle{rm}\Url}\fi

\bibitem[Ayala et~al.(2008)Ayala, Rosa, Wang, and Grabowski]{RN2}
O.~Ayala, B.~Rosa, L.-P. Wang, and W.~W. Grabowski.
\newblock Effects of turbulence on the geometric collision rate of sedimenting
  droplets. part 1. results from direct numerical simulation.
\newblock \emph{New Journal of Physics}, 10\penalty0 (7):\penalty0 075015,
  2008.

\bibitem[Ayala et~al.(2014)Ayala, Parishani, Chen, Rosa, and Wang]{RN7}
O.~Ayala, H.~Parishani, L.~Chen, B.~Rosa, and L.-P. Wang.
\newblock Dns of hydrodynamically interacting droplets in turbulent clouds:
  Parallel implementation and scalability analysis using 2d domain
  decomposition.
\newblock \emph{Computer Physics Communications}, 185\penalty0 (12):\penalty0
  3269--3290, 2014.

\bibitem[Batchelor and Green(1972)]{batchelor1972hydrodynamic}
G.~Batchelor and J.-T. Green.
\newblock The hydrodynamic interaction of two small freely-moving spheres in a
  linear flow field.
\newblock \emph{Journal of Fluid Mechanics}, 56\penalty0 (2):\penalty0
  375--400, 1972.

\bibitem[Bewley et~al.(2013)Bewley, Saw, and Bodenschatz]{RN5}
G.~P. Bewley, E.-W. Saw, and E.~Bodenschatz.
\newblock Observation of the sling effect.
\newblock \emph{New Journal of Physics}, 15\penalty0 (8):\penalty0 083051,
  2013.

\bibitem[Bordás et~al.(2013)Bordás, Roloff, Thévenin, and Shaw]{RN9}
R.~Bordás, C.~Roloff, D.~Thévenin, and R.~Shaw.
\newblock Experimental determination of droplet collision rates in turbulence.
\newblock \emph{New Journal of Physics}, 15\penalty0 (4):\penalty0 045010,
  2013.

\bibitem[Bragg and Collins(2014)]{RN6}
A.~D. Bragg and L.~R. Collins.
\newblock New insights from comparing statistical theories for inertial
  particles in turbulence: Ii. relative velocities.
\newblock \emph{New Journal of Physics}, 16, 2014.

\bibitem[Brunk et~al.(1997)Brunk, Koch, and Lion]{RN25}
B.~K. Brunk, D.~L. Koch, and L.~W. Lion.
\newblock Hydrodynamic pair diffusion in isotropic random velocity fields with
  application to turbulent coagulation.
\newblock \emph{Physics of Fluids}, 9\penalty0 (9):\penalty0 2670--2691, 1997.

\bibitem[Brunk et~al.(1998)Brunk, Koch, and Lion]{brunk1998turbulent}
B.~K. Brunk, D.~L. Koch, and L.~W. Lion.
\newblock Turbulent coagulation of colloidal particles.
\newblock \emph{Journal of Fluid Mechanics}, 364:\penalty0 81--113, 1998.

\bibitem[Cao et~al.(2008)Cao, Pan, de~Jong, Woodward, and Meng]{RN13}
L.~Cao, G.~Pan, J.~de~Jong, S.~Woodward, and H.~Meng.
\newblock Hybrid digital holographic imaging system for three-dimensional dense
  particle field measurement.
\newblock \emph{Applied Optics}, 47\penalty0 (25):\penalty0 4501--4508, 2008.

\bibitem[de~Jong et~al.(2010)de~Jong, Salazar, Woodward, Collins, and
  Meng]{RN11}
J.~de~Jong, J.~P. L.~C. Salazar, S.~H. Woodward, L.~R. Collins, and H.~Meng.
\newblock Measurement of inertial particle clustering and relative velocity
  statistics in isotropic turbulence using holographic imaging.
\newblock \emph{International Journal of Multiphase Flow}, 36\penalty0
  (4):\penalty0 324--332, 2010.

\bibitem[Dhariwal and Bragg(2018)]{RN26}
R.~Dhariwal and A.~D. Bragg.
\newblock Small-scale dynamics of settling, bidisperse particles in turbulence.
\newblock \emph{Journal of Fluid Mechanics}, 839:\penalty0 594--620, 2018.

\bibitem[Dou et~al.(2018{\natexlab{a}})Dou, Bragg, Hammond, Liang, Collins, and
  Meng]{RN16}
Z.~Dou, A.~D. Bragg, A.~L. Hammond, Z.~Liang, L.~R. Collins, and H.~Meng.
\newblock Effects of reynolds number and stokes number on particle-pair
  relative velocity in isotropic turbulence: a systematic experimental study.
\newblock \emph{Journal of Fluid Mechanics}, 839:\penalty0 271--292,
  2018{\natexlab{a}}.

\bibitem[Dou et~al.(2018{\natexlab{b}})Dou, Ireland, Bragg, Liang, Collins, and
  Meng]{RN17}
Z.~Dou, P.~J. Ireland, A.~D. Bragg, Z.~Liang, L.~R. Collins, and H.~Meng.
\newblock Particle-pair relative velocity measurement in high-reynolds-number
  homogeneous and isotropic turbulence using 4-frame particle tracking
  velocimetry.
\newblock \emph{Experiments in Fluids}, 59\penalty0 (2):\penalty0 30,
  2018{\natexlab{b}}.

\bibitem[Dou et~al.(2016)Dou, Pecenak, Cao, Woodward, Liang, and Meng]{RN20}
Z.~W. Dou, Z.~K. Pecenak, L.~J. Cao, S.~H. Woodward, Z.~Liang, and H.~Meng.
\newblock Piv measurement of high-reynolds-number homogeneous and isotropic
  turbulence in an enclosed flow apparatus with fan agitation.
\newblock \emph{Measurement Science and Technology}, 27\penalty0 (3), 2016.

\bibitem[Falkovich and Pumir(2007)]{falkovich2007sling}
G.~Falkovich and A.~Pumir.
\newblock Sling effect in collisions of water droplets in turbulent clouds.
\newblock \emph{Journal of the Atmospheric Sciences}, 64\penalty0
  (12):\penalty0 4497--4505, 2007.

\bibitem[Falkovich et~al.(2002)Falkovich, Fouxon, and
  Stepanov]{falkovich2002acceleration}
G.~Falkovich, A.~Fouxon, and M.~Stepanov.
\newblock Acceleration of rain initiation by cloud turbulence.
\newblock \emph{Nature}, 419\penalty0 (6903):\penalty0 151--154, 2002.

\bibitem[Ireland et~al.(2016)Ireland, Bragg, and Collins]{RN4}
P.~J. Ireland, A.~D. Bragg, and L.~R. Collins.
\newblock The effect of reynolds number on inertial particle dynamics in
  isotropic turbulence. part 1. simulations without gravitational effects.
\newblock \emph{Journal of Fluid Mechanics}, 796:\penalty0 617--658, 2016.

\bibitem[Kearney and Bewley(2020)]{RN10}
R.~V. Kearney and G.~P. Bewley.
\newblock Lagrangian tracking of colliding droplets.
\newblock \emph{Experiments in Fluids}, 61\penalty0 (7), 2020.

\bibitem[Larsen and Shaw(2018)]{RN23}
M.~L. Larsen and R.~Shaw.
\newblock A method for computing the three-dimensional radial distribution
  function of cloud particles from holographic images.
\newblock \emph{Atmospheric Measurement Techniques}, 11\penalty0 (7):\penalty0
  4261, 2018.

\bibitem[Lu and Shaw(2015)]{lu2015charged}
J.~Lu and R.~A. Shaw.
\newblock Charged particle dynamics in turbulence: Theory and direct numerical
  simulations.
\newblock \emph{Physics of Fluids}, 27\penalty0 (6):\penalty0 065111, 2015.

\bibitem[Meng et~al.(2004)Meng, Gang, Ye, and Woodward]{RN12}
H.~Meng, P.~Gang, P.~Ye, and S.~H. Woodward.
\newblock Holographic particle image velocimetry: from film to digital
  recording.
\newblock \emph{Measurement Science and Technology}, 15\penalty0 (4):\penalty0
  673, 2004.

\bibitem[Moffat(1988)]{RN27}
R.~J. Moffat.
\newblock Describing the uncertainties in experimental results.
\newblock \emph{Experimental Thermal and Fluid Science}, 1\penalty0
  (1):\penalty0 3--17, 1988.

\bibitem[Novara et~al.(2019)Novara, Schanz, Geisler, Gesemann, Voss, and
  Schröder]{RN19}
M.~Novara, D.~Schanz, R.~Geisler, S.~Gesemann, C.~Voss, and A.~Schröder.
\newblock Multi-exposed recordings for 3d lagrangian particle tracking with
  multi-pulse shake-the-box.
\newblock \emph{Experiments in Fluids}, 60\penalty0 (3):\penalty0 44, 2019.

\bibitem[Reade and Collins(2000)]{RN24}
W.~C. Reade and L.~R. Collins.
\newblock Effect of preferential concentration on turbulent collision rates.
\newblock \emph{Physics of Fluids}, 12\penalty0 (10):\penalty0 2530--2540,
  2000.

\bibitem[Rosa et~al.(2013)Rosa, Parishani, Ayala, Grabowski, and
  Wang]{rosa2013kinematic}
B.~Rosa, H.~Parishani, O.~Ayala, W.~W. Grabowski, and L.-P. Wang.
\newblock Kinematic and dynamic collision statistics of cloud droplets from
  high-resolution simulations.
\newblock \emph{New Journal of Physics}, 15\penalty0 (4):\penalty0 045032,
  2013.

\bibitem[Salazar et~al.(2008)Salazar, De~Jong, Cao, Woodward, Meng, and
  Collins]{RN14}
J.~P. L.~C. Salazar, J.~De~Jong, L.~J. Cao, S.~H. Woodward, H.~Meng, and L.~R.
  Collins.
\newblock Experimental and numerical investigation of inertial particle
  clustering in isotropic turbulence.
\newblock \emph{Journal of Fluid Mechanics}, 600:\penalty0 245--256, 2008.

\bibitem[Saw et~al.(2012{\natexlab{a}})Saw, Salazar, Collins, and
  Shaw]{saw2012spatial1}
E.-W. Saw, J.~P. Salazar, L.~R. Collins, and R.~A. Shaw.
\newblock Spatial clustering of polydisperse inertial particles in turbulence:
  I. comparing simulation with theory.
\newblock \emph{New Journal of Physics}, 14\penalty0 (10):\penalty0 105030,
  2012{\natexlab{a}}.

\bibitem[Saw et~al.(2012{\natexlab{b}})Saw, Shaw, Salazar, and
  Collins]{saw2012spatial2}
E.-W. Saw, R.~A. Shaw, J.~P. Salazar, and L.~R. Collins.
\newblock Spatial clustering of polydisperse inertial particles in turbulence:
  Ii. comparing simulation with experiment.
\newblock \emph{New Journal of Physics}, 14\penalty0 (10):\penalty0 105031,
  2012{\natexlab{b}}.

\bibitem[Saw et~al.(2014)Saw, Bewley, Bodenschatz, Ray, and Bec]{RN15}
E.-W. Saw, G.~P. Bewley, E.~Bodenschatz, S.~S. Ray, and J.~Bec.
\newblock Extreme fluctuations of the relative velocities between droplets in
  turbulent airflow.
\newblock \emph{Physics of Fluids (1994-present)}, 26\penalty0 (11):\penalty0
  111702, 2014.

\bibitem[Schanz et~al.(2012)Schanz, Gesemann, Schr{\"o}der, Wieneke, and
  Novara]{schanz2012non}
D.~Schanz, S.~Gesemann, A.~Schr{\"o}der, B.~Wieneke, and M.~Novara.
\newblock Non-uniform optical transfer functions in particle imaging:
  calibration and application to tomographic reconstruction.
\newblock \emph{Measurement Science and Technology}, 24\penalty0 (2):\penalty0
  024009, 2012.

\bibitem[Schanz et~al.(2016)Schanz, Gesemann, and Schröder]{RN21}
D.~Schanz, S.~Gesemann, and A.~Schröder.
\newblock Shake-the-box: Lagrangian particle tracking at high particle image
  densities.
\newblock \emph{Experiments in Fluids}, 57\penalty0 (5):\penalty0 70, 2016.

\bibitem[Sellappan et~al.(2020)Sellappan, Alvi, and
  Cattafesta]{sellappan2020lagrangian}
P.~Sellappan, F.~S. Alvi, and L.~N. Cattafesta.
\newblock Lagrangian and eulerian measurements in high-speed jets using
  multi-pulse shake-the-box and fine scale reconstruction (vic\#).
\newblock \emph{Experiments in Fluids}, 61\penalty0 (7):\penalty0 1--17, 2020.

\bibitem[Shaw(2003)]{RN1}
R.~A. Shaw.
\newblock Particle-turbulence interations in atmospheric clouds.
\newblock \emph{Annual Review of Fluid Mechanics}, 35\penalty0 (1):\penalty0
  183--227, 2003.

\bibitem[Squires and Eaton(1991)]{squires1991preferential}
K.~D. Squires and J.~K. Eaton.
\newblock Preferential concentration of particles by turbulence.
\newblock \emph{Physics of Fluids A: Fluid Dynamics}, 3\penalty0 (5):\penalty0
  1169--1178, 1991.

\bibitem[Sundaram and Collins(1997)]{RN3}
S.~Sundaram and L.~R. Collins.
\newblock Collision statistics in an isotropic particle-laden turbulent
  suspension .1. direct numerical simulations.
\newblock \emph{Journal of Fluid Mechanics}, 335:\penalty0 75--109, 1997.

\bibitem[Tavoularis et~al.(1978)Tavoularis, Bennett, and
  Corrsin]{tavoularis1978velocity}
S.~Tavoularis, J.~Bennett, and S.~Corrsin.
\newblock Velocity-derivative skewness in small reynolds number, nearly
  isotropic turbulence.
\newblock \emph{Journal of Fluid Mechanics}, 88\penalty0 (1):\penalty0 63--69,
  1978.

\bibitem[Wang et~al.(2000)Wang, Wexler, and Zhou]{wang2000statistical}
L.-P. Wang, A.~S. Wexler, and Y.~Zhou.
\newblock Statistical mechanical description and modelling of turbulent
  collision of inertial particles.
\newblock \emph{Journal of Fluid Mechanics}, 415:\penalty0 117--153, 2000.

\bibitem[Wang et~al.(2005)Wang, Ayala, Kasprzak, and Grabowski]{RN8}
L.-P. Wang, O.~Ayala, S.~E. Kasprzak, and W.~W. Grabowski.
\newblock Theoretical formulation of collision rate and collision efficiency of
  hydrodynamically interacting cloud droplets in turbulent atmosphere.
\newblock \emph{Journal of the Atmospheric Sciences}, 62\penalty0 (7):\penalty0
  2433--2450, 2005.

\bibitem[Wang et~al.(2008)Wang, Ayala, Rosa, and Grabowski]{wang2008turbulent}
L.-P. Wang, O.~Ayala, B.~Rosa, and W.~W. Grabowski.
\newblock Turbulent collision efficiency of heavy particles relevant to cloud
  droplets.
\newblock \emph{New Journal of Physics}, 10\penalty0 (7):\penalty0 075013,
  2008.

\bibitem[Wieneke(2008)]{RN22}
B.~Wieneke.
\newblock Volume self-calibration for 3d particle image velocimetry.
\newblock \emph{Experiments in Fluids}, 45\penalty0 (4):\penalty0 549--556,
  2008.

\bibitem[Yavuz et~al.(2018)Yavuz, Kunnen, van Heijst, and Clercx]{RN18}
M.~A. Yavuz, R.~P.~J. Kunnen, G.~J.~F. van Heijst, and H.~J.~H. Clercx.
\newblock Extreme small-scale clustering of droplets in turbulence driven by
  hydrodynamic interactions.
\newblock \emph{Physical Review Letters}, 120\penalty0 (24):\penalty0 244504,
  2018.

\bibitem[Zhou et~al.(2001)Zhou, Wexler, and Wang]{zhou2001modelling}
Y.~Zhou, A.~S. Wexler, and L.-P. Wang.
\newblock Modelling turbulent collision of bidisperse inertial particles.
\newblock \emph{Journal of Fluid Mechanics}, 433:\penalty0 77, 2001.

\end{thebibliography}

\end{document}